\documentclass[floats,preprint,prb,aps]{revtex4}
\begin{document}
\title{Current Profiles of Molecular Nanowires; DFT Green Function Representation}
\author{Oleg Berman and Shaul Mukamel}
\affiliation{Department of Chemistry, University of California
Irvine, California 92697-2025}

\vspace{0.5cm}

\preprint{\em Submitted to {\it  Phys. Rev.} {\bf B} }

\date{\today}
\vspace{2.7in}

\begin{abstract}

The Liouville-space Green function formalism is used to compute
the current density profile across a single molecule attached to
electrodes. Time ordering is maintained in real, physical, time,
avoiding the use of artificial time loops and backward
propagations. Closed expressions for molecular currents, which
only require DFT calculations for the isolated molecule, are
derived to fourth order in the molecule/electrode coupling.

\vspace{1.0 in}



\end{abstract}

\maketitle

\section{Introduction}

There is a considerable interest in
measuring~\cite{Lee}$^-$~\cite{Cui} and
computing~\cite{Xue2}$^-$\cite{Kosov}$^,$~\cite{Mujica1}$^-$~\cite{Lang_Avouris}
currents in single molecules attached to electrodes. The
current-voltage ($I-V$) characteristics of such devices are
strongly nonlinear, and should be recast in terms of high order
conductivities~\cite{issue,Reed,Cui}. Controlling molecular
currents has important applications to nanodevices such as
molecular wires and rectifiers~\cite{issue,Reed,Cui}.

The first-principles computation of molecular currents poses a
serious challenge that was addressed by many
approaches~\cite{Xue,Kosov}. Standard quantum chemistry packages
are not designed to compute current-carrying states.
Non-equilibrium Green function techniques widely used for
computing currents in macroscopic systems, e.g.
semiconductors~\cite{Datta,Haug}, were adopted for single-molecule
currents~\cite{Xue2,Xue1,Xue}. The underlying diagrammatic
perturbative expansion requires the computation of extra
information in the intermediate steps, which is not needed for
computing the current. This includes the dependence of the Green
function on two times, involving coherence between states with a
different number of electrons~\cite{Xue}. In a different approach
the time dependent charge distribution in a donor-acceptor
junction was studied by calculating the tunnelling currents at the
Hartree-Fock level in terms of the donor and acceptor
states~\cite{Stuchebrukhov}.

An ab initio algorithm for calculating molecular currents
developed by Lang~\cite{Lang,Lang_Avouris} is based on the
self-consistent solution of a system of four  equations:  (1) The
Kohn-Sham integral equation for the wavefunction of the molecule +
electrodes (Eq.~(3.1) in Ref.\cite{Lang}), (2) Equation for the
potential difference between the complete (molecule + electrodes)
system and the bare electrodes (Eq.~(3.2) in Ref.\cite{Lang}), (3)
Equation for the Green function of the molecule in an effective
potential given by the sum of the exchange-correlation and the
electrostatic potential (Eq.~(3.13) in Ref.\cite{Lang}). The
latter is determined by (4) the Poisson equation with boundary
conditions setting the potential to be equal to the chemical
potentials difference of the electrodes (Eq.~(\ref{peq0}) in this
paper). These equations, together with the boundary conditions,
yield  the wavefunction of the entire (molecule + electrodes)
system which can be used to compute the finite current. This
algorithm was applied for calculating the current in a chain of
three  $Al$ atoms  with the substitution of one atom by sulfur,
and for a chain of ten carbon atoms. The electrodes were modelled
as a semi-infinite ideal metal (homogeneous electron gas or
Jellium model). The same approach was subsequently applied to
calculate the current in a benzene-$1,4$-dithiol molecule
connected to ideal metal electrodes, and the effect of inserting a
single gold atom between the molecule and a metallic surface was
investigated~\cite{diventra}.

Mujica, Kemp, Roitberg and Ratner~\cite{Mujica} proposed to solve
self-consistently the Hartree-Fock equation for the Green function
of the complete  (molecule + electrodes) system and the Poisson
equation for the electrostatic potential, with the same boundary
conditions of Lang~\cite{Lang,Lang_Avouris}. The molecule was
described by the Hubbard Hamiltonian, the electrodes were taken to
be an ideal metal,  and the current is computed from the Green
function. This approach is easier to implement compared to the
density functional theory (DFT)~\cite{Lang,Lang_Avouris}, since it
only requires the solution of two rather than four self-consistent
equations. However, DFT~\cite{Lang,Lang_Avouris} takes into
account exchange-correlation effects more rigorously.

Kosov applied a variational Kohn-Sham density-functional (DFT)
equation for electrons in molecular wires
(benzene-$1,4$-dithiolate molecule covalently bonded to two Gold
electrodes)~\cite{Kosov}. By variational minimization of the total
energy, where an arbitrary given current is imposed as a
constraint~\cite{Kosov}, a ten atom Gold cluster participates in
the bonding with a molecule. The device contains $32$ atoms. This
results in a system of two self-consistent equations, similar to
the Hartree-Fock approach~\cite{Mujica}, which is simpler than the
equations of Ref.~\cite{Lang,Lang_Avouris}, yet treats
exchange-correlation effects as rigorously as
DFT~\cite{Lang,Lang_Avouris}.

All of these approaches, which employ DFT for the molecule +
electrodes, require the solution of a system of self-consistent
equations, but do not give closed expressions for the current.

In this paper we compute the current profile in a molecule
perturbatively in the ratio of the coupling of the molecular
electrons with the electrodes and the separation between Kohn-Sham
orbital energies. Our approach yields a closed expression for the
current density $\left \langle \hat{J}(\mathbf{r}) \right \rangle$
in terms of  Kohn-Sham orbitals of the molecule alone, which enter
various Green functions in the frequency domain. The
molecule/electrodes coupling potential of Lang's
approach~\cite{Lang} is treated to the lowest order required to
yield a finite current.  We use a Liouville space (superoperator)
many-body Green function technique, which provides a convenient
description of non-equilibrium effects~\cite{Muk_sup,Mukamel}. The
Liouville space TDDFT (Time-Dependent Density Functional Theory)
formulation of non-linear response based on the single electron
density matrix was developed in Ref.~\cite{Berman,Tretiak}, and
subsequently extended to superconductors~\cite{Berman_sup}.  One
advantage of working in the higher dimensional Liouville space is
that we need only consider time ordered quantities in real
(physical) time and Wick's theorem therefore assumes a
particularly compact form~\cite{Muk_sup}; in contrast to Hilbert
space Green function perturbation theory~\cite{kp,Keldysh,Haug} no
special contours or analytic continuations are necessary.

\section{The Model Hamiltonian}

We shall partition the electronic Hamiltonian of  a molecule
connected to two electrodes $A$ and $B$, in the form
\begin{eqnarray}
\label{ham} \hat{H} &=& \hat{H}_{0} + \hat{V},
\end{eqnarray}
where $\hat{H}_{0}$ represents the noninteracting molecule and
electrodes, and $\hat{V}$ is the molecule/electrodes coupling
\begin{eqnarray}
\label{V} \hat{V} = \int dx \int dx' U(x,x')\left[
\psi^{\dagger}(x) (\psi^{A}(x') + \psi^{B}(x')) +
(\psi^{A\dagger}(x') + \psi^{B\dagger}(x'))\psi(x) \right].
\end{eqnarray}
The spatial coordinates $x$ and $x'$ run over the molecule and the
electrode  regions, respectively ($x \equiv (\mathbf{r},t)$ is a
coordinate of electrons within the molecule, and $x' \equiv
(\mathbf{r}',t)$ is a coordinate of electrons in the electrodes
region).  $U(x,x')$ is the coupling potential  between the
molecule and electrodes; $\psi(x)$ is the field operator of
electrons in the molecule, whereas $\psi^{A}(x')$ and
$\psi^{B}(x')$ are the field operators of electrons in the two
electrodes. These operators satisfy Fermi anti-commutation
relations~\cite{Abrikosov,Fetter}
\begin{eqnarray}
\label{fermicom} && \psi({\bf r}_{1})\psi^{\dagger}({\bf r}_{2}) +
\psi^{\dagger}({\bf r}_{2}) \psi({\bf r}_{1}) = \delta
(\mathbf{r}_{1} - \mathbf{r}_{2}); \nonumber
\\ && \psi({\bf r}_{1})\psi({\bf r}_{2}) + \psi({\bf
r}_{2}) \psi({\bf r}_{1}) = 0 ; \nonumber \\ &&
\psi^{\dagger}({\bf r}_{1})\psi^{\dagger}({\bf r}_{2}) +
\psi^{\dagger}({\bf r}_{2}) \psi^{\dagger}({\bf r}_{1}) = 0 .
\end{eqnarray}

 We assume that the electrodes interact with the molecule only
at two points of contact ($\mathbf{r_{0}^{A}}$ and
$\mathbf{r_{0}^{B}}$). Kosov considered the coupling between the
molecule and the electrodes represented by $10$ Gold
atoms~\cite{Kosov}; participation of only a restricted domain of
the electrodes in the molecule/electrode coupling was also
assumed in the ``extended molecule'' model~\cite{Xue}. We then
have
\begin{eqnarray}
\label{loc}   U (\mathbf{r}, \mathbf{r}') = \phi (\mathbf{r}) [
\delta (\mathbf{r}' - \mathbf{r_{0}^{A}}) +  \delta (\mathbf{r}' -
\mathbf{r_{0}^{B}})] ,
\end{eqnarray}
where $\phi (\mathbf{r})$ is defined as the potential, induced by
the difference  between the charge density of the isolated
molecule $\varrho_{0}(\mathbf{r})$ and of the molecule coupled to
the electrodes $\varrho(\mathbf{r})$~\cite{Xue}.   $\phi
(\mathbf{r})$ is obtained by solving the Poisson equation in the
molecule ($r_{0}^{A} < r < r_{0}^{B}$) (Eq.~(5.3) in
Ref.~\cite{Xue})
\begin{eqnarray}
\label{varloc}  \nabla^{2}  \phi (\mathbf{r}) = - e (\varrho
(\mathbf{r}) - \varrho_{0}(\mathbf{r}))  ,
\end{eqnarray}
with the boundary conditions  $\phi(\mathbf{r_{0}^{A}}) = \mu_{A}$
and $\phi(\mathbf{r_{0}^{B}}) = \mu_{B}$. Here $\mu_{A}$ and
$\mu_{B}$ are the chemical potentials of electrodes $A$ and $B$,
respectively, $W = \mu_{A} - \mu_{B}$ is the external voltage, and
$e$ is electron charge.

The Poisson equation  for the total electrostatic potential in the
molecule coupled to the electrodes  $U_{el}(\mathbf{r})$
is~\cite{Xue,Mujica1}
\begin{eqnarray}
\label{peq0} \nabla^{2}  U_{el}(\mathbf{r}) = - e \varrho
(\mathbf{ r})  ,
\end{eqnarray}
with the boundary conditions $U_{el}(\mathbf{r_{0}^{A}}) =
\mu_{A}$ and $U_{el}(\mathbf{r_{0}^{B}}) = \mu_{B}$. In act the
boundary conditions should reflect the change of the electrostatic
potential in the interior of the left (right) electrodes when
current is flowing, rather than the chemical potential difference
of electrodes  (Eq.~(5.4) in Ref.~\cite{Xue}). The latter between
the last two is, in general, smaller than the applied
voltage\cite{Lang1}. This difference should be negligible for
large electrode spacings, for which the current is small, but
should be noticeable for small spacings. Our boundary conditions
thus hold in the small current limit (large electrode spacings).

To lowest order in the molecule/electrodes coupling we use  the
unperturbed charge density of the isolated molecule in the r.h.s.
of the Poisson equation $\varrho (\mathbf{ r}) =  \varrho_{0}
(\mathbf{ r})$ , and Eq.~(\ref{peq0}) gives
\begin{eqnarray}
\label{peq} \nabla^{2}  U_{el}(\mathbf{r}) = - e  \sum_{\alpha}
\varphi_{\alpha}(\mathbf{r})\varphi_{\alpha}^{*}(\mathbf{ r})  ,
\end{eqnarray}
where the sum runs over all occupied molecular Kohn-Sham orbitals
 $\varphi_{\alpha }(\mathbf{r})$.  Using $\varrho (\mathbf{ r}) =  \varrho_{0} (\mathbf{
 r})$, we
obtain from Eq.~(\ref{varloc}) (Eq.~(5.6) in Ref.~\cite{Xue})
$\nabla^{2} \phi (\mathbf{r}) = 0$, $\phi (\mathbf{r_{0}^{A}}) =
\mu_{A}$, $\phi (\mathbf{r_{0}^{B}}) = \mu_{B}$. By solving this
equation, we obtain that the coupling potential is linear in
$\mathbf{r}$
\begin{eqnarray}
\label{peq2}  \phi (\mathbf{r}) = (\mathbf{r_{0}^{A}} -
\mathbf{r_{0}^{B}})^{-1}\left(\mu_{B}\mathbf{r_{0}^{A}} -
\mu_{A}\mathbf{r_{0}^{B}} + (\mu_{A} - \mu_{B})\mathbf{r}\right).
\end{eqnarray}
There are two contributions to the charge density in the Poisson
equation (Eq.~(\ref{peq0})): The charge density of the isolated
molecule and the variation of the charge density with the
electrodes due to the coupling  induced by the external
voltage~\cite{Lang,Xue}. To lowest order in the coupling potential
we neglect the latter. And the electrostatic potential in the
molecule is calculated using the unperturbed charge density
(Eq.~(\ref{peq})). We thus neglect all screening effects and the
potential profile across the junction is assumed to be linear
(Eq.~(\ref{peq2})). This holds for low voltages. The electronic
structure of the molecule enters in $\hat{H}_{0}$. The external
voltage, which is the difference between the chemical potentials
of the electrodes ($W \equiv \mu_{A} - \mu_{B}$), enters in the
Hamiltonian of the molecule/electrodes coupling $\hat{V}$ in
Eq.~(\ref{ham}) through the coupling potential $U$
(Eqs.~(\ref{loc}) and~ (\ref{peq2})). Our perturbative approach
holds for low voltages, where provided the coupling potential of
the molecular electrons with the electrodes (the external voltage
(Eq.~(\ref{peq2}))) is much smaller than the difference between
Kohn-Sham orbital energies.

It corresponds to lowest order expansion in the
molecule/electrodes coupling potential of Lang's
approach~\cite{Lang}.

\section{Perturbative calculation of the current}

Our goal is to compute the expectation value of the current
operator
\begin{eqnarray}
\label{cur}\hat{J}(x) = - \left. \frac{ie\hbar}{2m}
\left(\nabla_{\mathbf{r}} - \nabla_{\mathbf{\bar{r}}}\right)
\psi(x)\psi^{\dagger}(\bar{x})\right|_{x = \bar{x}} ,
\end{eqnarray}
where $e$ and $m$ are the electron charge and mass, respectively.

We start with the interaction picture expression for the
expectation value of $\hat{J}(x)$, using the partitioning
(Eq.~(\ref{ham})) of the Hamiltonian~\cite{Muk_sup}. We define the
Liouville operators $ L = L_{0} + V_{-}$ corresponding to
Eq.~(\ref{ham}) where $L_0 \equiv (H_0)_{-}$ i.e., $L_0 X \equiv
H_0 X -XH_0$. The zero order time evolution operator is defined as
\begin{equation}
\label{4} \mathcal{G}_{0}(\tau_{2},\tau_{1}) \equiv \theta
(\tau_{2}-
\tau_{1})\exp\left[-\frac{i}{\hbar}L_{0}(\tau_{2}-\tau_{1})\right],
\end{equation}
where $\theta(t)$ is the Heavyside step function. Throughout this
paper we denote  operators in the interaction picture by a tilde
(~$\tilde{}$~), i.e.
\begin{eqnarray}
\label{7}
 \tilde{A}_\nu (\tau)\equiv \mathcal{G}_{0}^{\dag} (\tau, 0)
A_{\nu} \mathcal{G}_0(\tau, 0)&\\\nonumber &\nu = +,-,\,\, or \,\,
\nu = L,R.
\end{eqnarray}

The current is given by
\begin{eqnarray}
\label{excur}\left \langle \hat{J}(x) \right \rangle = - \left
\langle \hat{T} \tilde{J}_{+}(x)exp\left[- \frac{i}{\hbar}
\int_{-\infty}^{t} d\tau \tilde{V}_{-}(\tau)\right] \right \rangle
,
\end{eqnarray}
where the Liouville space superoperators with indices $+$ and $-$
 are defined in Appendix~\ref{ap.1pr}~\cite{Muk_sup}, and the
time-ordering operator in Liouville space  $\hat{T}$ is defined in
(Eq.~(\ref{3}))~\cite{Muk_sup}. This natural time ordering
chronologically follows the various interactions with the
electrodes. The precise order in which superoperators appear next
to a $\hat{T}$ operator is immaterial since at the end the order
will be fixed by $\hat{T}$. $\hat{T}$ before an exponent means
that each term in the Taylor expansion of this exponent should be
time-ordered.

The current can be also  expressed in terms of the molecular
electron density matrix $\tilde{\rho}(\mathbf{r},\mathbf{r}',t)$
in the interaction picture
\begin{eqnarray}
\label{rho}\tilde{\rho}(\mathbf{r}, \mathbf{\bar{r}}, t) \equiv
Tr\left[\tilde{\psi}(\mathbf{r},t)
\tilde{\psi}^{\dagger}(\bar{\mathbf{r}},t)\rho_{0}\right] ,
\end{eqnarray}
where $\rho_{0}$ is unperturbed many-electron density matrix of
the isolated molecule uncoupled to the electrodes. The expression
for the current is
\begin{eqnarray}
\label{currho} \left \langle \hat{J}(x) \right \rangle = - \left.
\frac{ie\hbar}{2m} \left(\nabla_{\mathbf{r}} -
\nabla_{\mathbf{\bar{r}}}\right) \tilde{\rho} (\mathbf{r},
\mathbf{\bar{r}},t)\right|_{x = \bar{x}} .
\end{eqnarray}

A finite current requires at least two interactions with each
electrode. We, therefore, computed the current to fourth order in
$\tilde{V}$
\begin{eqnarray}
\label{pertexp} \left \langle \hat{J}(x) \right \rangle = - \left.
\frac{ie\hbar}{2m}\left(\nabla_{\bar{\mathbf{r}}} -
\nabla_{\mathbf{r}}\right) S(x,\bar{x})\right|_{x = \bar{x}} ,
\end{eqnarray}
where
\begin{eqnarray}
\label{sx} S(x,\bar{x}) &=& - \frac{1}{4!}\left(
\frac{i}{\hbar}\right)^{4}\int_{-\infty}^{t}
d\tau_{1}\int_{-\infty}^{t} d\tau_{2}\int_{-\infty}^{t} d\tau_{3
}\int_{-\infty}^{t} d\tau_{4}  \nonumber
\\ && \int d\mathbf{r}_{1} \int
d\mathbf{r'}_{1}\int d\mathbf{r}_{2}\int d\mathbf{r'}_{2}\int
d\mathbf{r}_{3}\int d\mathbf{r'}_{3}\int d\mathbf{r}_{4}\int
d\mathbf{r'}_{4} \nonumber
\\ &&\left\langle
\hat{T}\tilde{\rho}(x,\bar{x},t)\tilde{V}_{-}(\mathbf{r}_{4},\tau_{4})
\tilde{V}_{-}(\mathbf{r}_{3},\tau_{3})
\tilde{V}_{-}(\mathbf{r}_{2},\tau_{2})\tilde{V}_{-}(\mathbf{r}_{1},\tau_{1})
\right\rangle _{0} ,
\end{eqnarray}
where $\tilde{\rho}(x,\bar{x},t) =
\tilde{\psi}(x)\tilde{\psi}^{\dagger}(\bar{x})$.

We next expand $S(x,\bar{x})$ in Eq.~(\ref{sx}) in ``left'' and
``right'' operators in Liouville space~\cite{Muk_sup}. To that end
we introduce the field operator corresponding to both electrodes
\begin{eqnarray}
\label{psiel} \psi^{M}(x') \equiv \psi^{A}(x') + \psi^{B}(x') .
\end{eqnarray}
Substituting Eq.~(\ref{V}) into Eq.~(\ref{psiel}), we get
\begin{eqnarray}
\label{sbig} S(x,\bar{x}) &=& \frac{1}{4!}\left(
\frac{i}{\hbar}\right)^{4}\int_{-\infty}^{t}
d\tau_{1}\int_{-\infty}^{t} d\tau_{2}\int_{-\infty}^{t} d\tau_{3
}\int_{-\infty}^{t} d\tau_{4}  \nonumber
\\ && \int d\mathbf{r}_{1} \int
d\mathbf{r'}_{1}\int d\mathbf{r}_{2}\int d\mathbf{r'}_{2}\int
d\mathbf{r}_{3}\int d\mathbf{r'}_{3}\int d\mathbf{r}_{4}\int
d\mathbf{r'}_{4} \nonumber
\\ &&
U(x_{4},x_{4}^{'})U(x_{3},x_{3}^{'})U(x_{2},x_{2}^{'})U(x_{1},x_{1}^{'})I(x_{1},x_{1}
^{'}, x_{2},x_{2}^{'},x_{3},x_{3}^{'}, x_{4},x_{4}^{'},x,\bar{x})
,
\end{eqnarray}
where $I$ is given by Eq.~(\ref{if}). In Appendix \ref{ap.1pr} we
use superoperator algebra to bring it to the form of
Eq.~(\ref{if1}). Factorizing  that equation, using Wick's theorem
for fermion superoperators (Eq.~(74) in Ref.\cite{Muk_sup}), we
obtain
\begin{eqnarray}
\label{if2} I  &=& \frac{1}{4!} \sum_{\nu_{\eta},\nu_{\kappa} =
0,1; \  i,j  = 0,1} (-1)^{P} \left[ \prod_{\eta,\kappa = 0,\ldots,
4; \ k,l = 0,1}
 \langle
\hat{T}(\psi_{\nu_{\eta}}^{M_{k}\alpha_{i}}(x_{\eta}^{k})\psi_{\nu_{\kappa}}^{M_{l}\alpha_{j}}(x_{\kappa}^{l})
 )\rangle \right. \nonumber \\
  && \left.(1 - \delta_{\eta,0}\delta_{\nu_{\kappa},1}\delta_{k,1})
 (1 - \delta_{\kappa,0}\delta_{\nu_{\kappa},1}\delta_{k,1})
 \right],
\end{eqnarray}
where $\delta$ is the Kronecker symbol: for $\nu_{\eta} = 0$
$\psi_{\nu_{\eta}} = \psi_{L}$, and for $\nu_{\eta} = 1$
$\psi_{\nu_{\eta}} = \psi_{R}$; for $k = 1$ $\psi^{M_{k}}(x^{k}) =
\psi^{M}(x')$, for $k = 0$ $\psi^{M_{k}}(x^{k}) = \psi(x)$; for $i
= 0$ $\psi^{\alpha_{i}} = \psi$, and for $i = 1$
$\psi^{\alpha_{i}} = \psi^{\dagger}$. $P = 1/2\left(n_{L}^{dif} +
n_{R}^{dif}\right)$ is a number of sum of permutations of
$\psi_{L}$ and $\psi_{R}$, where $n_{L}^{dif}$ is a number of
contractions in the r.h.s. of Eq.~(\ref{if2}) $\langle \hat{T}
(\psi_{L}(x_{n})\psi_{L}(x_{m})) \rangle $, when $n \neq m$, and
$n_{R}^{dif}$ is a number of contractions in the r.h.s. of
Eq.~(\ref{if2}) $\langle \hat{T}
(\psi_{R}(x_{n})\psi_{R}(x_{m}))\rangle$, when $n \neq m$.

We assume that the molecule contacts each electrode at a single
point ($\mathbf{r_{0}^{A}}$ and $\mathbf{r_{0}^{B}}$).  The
coupling potential between the molecule and electrode is
\begin{eqnarray}
\label{potcoup} U(\mathbf{r},\mathbf{r}') = \sum_{\alpha \beta}
U_{\alpha \beta}\varphi_{\alpha
}^{*}(\mathbf{r})\varphi_{\beta}(\mathbf{r}') ,
\end{eqnarray}
where $U(\mathbf{r},\mathbf{r}')$ is determined by
Eq.~(\ref{loc}); $\varphi_{\alpha }(\mathbf{r})$ is the
$\alpha$'th Kohn-Sham orbital
 of the  molecule, and
\begin{eqnarray}
\label{uan} U_{\alpha \beta} = \int d\mathbf{r}\int d\mathbf{r}'
U(\mathbf{r},\mathbf{r}')\varphi_{\alpha}(\mathbf{r})\varphi_{\beta}^{*}(\mathbf{
r}') = (\varphi_{\beta}^{*}(\mathbf{r_{0}^{A}}) +
\varphi_{\beta}^{*}(\mathbf{r_{0}^{B}})) \int d\mathbf{r} \phi
(\mathbf{r})\varphi_{\alpha}(\mathbf{r}) ,
\end{eqnarray}
where $\phi (\mathbf{r})$ is determined by Eq.~(\ref{peq2}).
 Below we denote
$U_{\alpha } \equiv U_{\alpha \alpha} $.

Substituting the expansion of the  frequency domain Green
functions and coupling potential in Kohn-Sham orbitals
(Eqs.~(\ref{dft_green_defin}) and~(\ref{potcoup})) in the
expressions for $I$ (Appendices~\ref{ap.2pr} and~\ref{ap.3pr}),
and substituting $I$ in Eq.~(\ref{sbig}) and $S$ in
Eq.~(\ref{pertexp}), we obtain
\begin{eqnarray}
\label{current_res} \left \langle \hat{J}(x) \right \rangle &=&
-\frac{e}{m\hbar^{3}}\left. \left(\nabla_{\bar{\mathbf{r}}} -
\nabla_{\mathbf{r}}\right)\right|_{\mathbf{r} \rightarrow
\bar{\mathbf{r}}}\int
\frac{d\omega}{2\pi}\int\frac{d\omega_{1}}{2\pi}\int\frac{d\omega_{2}}{2\pi}\delta(\omega
- \omega_{1} - \omega_{2}) \nonumber\\ && \left[ \sum_{\beta
=
1}^{N}\sum_{i=1}^{21}I_{i}\varphi_{\beta}(\mathbf{r})\varphi_{\beta}^{*}(\bar{\mathbf
{r}}) \right].
\end{eqnarray}
Eq.~(\ref{current_res}) is our final expression for the current.
The twenty one terms $I_{i}$ contributing to the current  are
given in Appendix~\ref{ap.4pr}.

If the external voltage is zero, $\mu_{A} = \mu_{B} = \mu_{0}$.
Setting $\mu_{0} = 0$,  Eq.~(\ref{peq2}) gives $\phi(\mathbf{r}) =
0$ and $U(\mathbf{r},\mathbf{r}') = 0$ everywhere in the molecule.
All twenty one terms $I_{i}$ contributing to the current
 Eqs.~(\ref{ie1})-~(\ref{it9}) then vanish.

\section{Discussion}

In this paper we have derived a closed expression
(Eq.~(\ref{current_res}) together with Appendix~\ref{ap.4pr}) for
the current profile within a molecule attached to electrodes, in
terms of Green functions of the isolated molecule. The external
voltage $W \equiv \mu_{A} - \mu_{B}$ is equal to the difference
between the chemical potentials of  electrodes $A$ and $B$. Only
one characteristic of the electrodes, the Fermi energy levels
$\mu_{A}$ and $\mu_{B}$, enters their Green functions, the
molecule/electrodes coupling potential and the final expression
for the molecular current
(Eqs.~(\ref{dft_green_corfrm})-~(\ref{ggff1m})). This expression
only requires DFT calculations for the molecule  alone (which
enters in the Green functions of the molecule in the expression
for the current) rather than the molecule and electrodes as used
in~Refs.\cite{Xue}$^-$\cite{Kosov}$^,$\cite{Mujica}$^-$\cite{Lang_Avouris}.
The derivation employs the Liouville space Green function
technique~\cite{Muk_sup}, and the current is expanded to fourth
order in the molecule/electrodes coupling our formula for the
current (Eq.~(\ref{current_res})) should agree with Lang's
approach~\cite{Lang} to lowest order in the  molecule/electrodes
coupling potential. The closed expressions for the current
obtained in the paper, compared to the previous considerations, is
not the result of the new methodology, but stems from the
perturbative considerations.

We have established a one-to-one correspondence between Liouville
space and Hilbert space Green functions for fermions
(Appendix~\ref{ap.2pr} and Appendix~\ref{ap.3pr}). For
completeness, similar relations for Boson Green functions are
given in Appendix~\ref{ap.5pr}. The only difference is the
definition of the time-ordering in Liouville space (Eq.~(\ref{3})
for fermions {\it vs.} Eq.~(\ref{31}) for bosons).

Liouville space provides a fully time ordered description of the
current since we only need to propagate the density matrix forward
in time. In contrast, the wavefunction in Hilbert space involves
both forward and backward propagations. The choice is between
following the ket only, moving forward and backward or following
the joint dynamics of the ket and the bra and moving only
forward~\cite{Muk_sup}. Artificial time variables (Keldysh loops)
commonly used in many-body theory~\cite{kp,Keldysh,Haug}, which
are necessary for the wavefunction picture, are avoided by the
density matrix  in Liouville space which uses the real laboratory
timescale throughout the calculation.

\section*{Acknowledgements}

The support of the National Science Foundation grant no.
CHE-0132571 is gratefully acknowledged.

\appendix

==================================

\section{Superoperator Algebra for the Current}\label{ap.1pr}

A Liouville space operator $A_\alpha$ is labelled by a Greek
subscript where $\alpha =L, R, +, -$. It is defined by its action
on an ordinary (Hilbert space) operator  $X$~\cite{Muk_sup}.
\begin{equation}
\label{A1} (A_\alpha X)_{ij} \equiv \sum_{\kappa\ell}
(A_\alpha)_{i j, \kappa \ell} X_{\kappa\ell}
\end{equation}
$A_\alpha$ is thus a \emph{tetradic} operator with four indices.
$A_L X\equiv A X$ denotes action from the left, and $A_R X \equiv
X A$ denotes action from the right. We then obtain using
Eq.~(\ref{A1})
\begin{equation}
\label{A2} (A_L)_{ij, \kappa\ell} = A_{i\kappa} \delta_{j\ell}
\end{equation}
\begin{equation}
\label{A3}
 (A_R)_{ij, \kappa\ell} = A_{\ell j} \delta_{i\kappa}
\end{equation}
Note that the order of the $j\ell$ indices in Eq.~(\ref{A3}) has
been reversed.

We further define the symmetric and the antisymmetric combinations
$A_+ \equiv \frac{1}{2}(A_L + A_R)$ and $A_-\equiv A_L-A_R\,$. We
then have the tetrahedral matrix elements $(A_{-})_{ij,\kappa\ell}
= A_{i\kappa} \delta_{j\ell} - A_{\ell j} \delta_{i\kappa}$,
$(A_{+})_{ij,\kappa\ell} = \frac{1}{2}[A_{i\kappa} \delta_{j\ell}
+ A_{\ell j} \delta_{i\kappa}]$. Note that $[A_L, B_R]=0$. This
commutativity of left and right operators made possible thanks to
the larger size of Liouville space, simplifies algebraic
manipulations resulting in many useful relations\cite{Muk_sup}.

Using the above definitions of the Liouville space algebra, we can
represent $I$ in Eq.~(\ref{sbig}) in the form
\begin{eqnarray}
\label{if} I  &=& \frac{1}{4!} \sum_{\alpha \neq \beta =0,1}
 \langle \hat{T}\psi(x)
\psi^{\dagger}(\bar{x})(\psi^{\alpha}(x_{1})
\psi^{M\beta}(x_{1}^{'}))_{-} ( \psi^{\alpha}(x_{2})
\psi^{M\beta}(x_{2}^{'}))_{-}( \psi^{\alpha}(x_{3})
\psi^{M\beta}(x_{3}^{'}))_{-} \nonumber \\ &&(
\psi^{\alpha}(x_{4}) \psi^{M\beta}(x_{4}^{'}))_{-} \rangle .
\end{eqnarray}
Here, for $\alpha = 1$ $\psi^{\alpha}(x_{1}) =
\psi^{\dagger}(x_{1})$, and, for $\alpha = 0$
$\psi^{\alpha}(x_{1}) = \psi(x_{1})$. In the r.h.s. of
Eq.~(\ref{if}) there are sixteen terms.

We can now expand Eq.~(\ref{if}) in ``left'' and ``right''
Liouville space operators\cite{Muk_sup}. We note that the
superoperator corresponding to  any function $f$ of an operator
$Q_{j}$ can be expressed in terms of $Q_{j+}$ and $Q_{j-}$,
i.e.,\cite{Muk_sup}
\begin{eqnarray}
\label{63} [f(Q_{j})]_{-} \equiv f (Q_{jL}) - f (Q_{jR}) &=&
f\left(Q_{j+}+ \frac{1}{2} Q_{j-}\right) -
f\left(Q_{j+}-\frac{1}{2} Q_{j-}\right),
\end{eqnarray}
and
\begin{eqnarray}
\label{63b} 2\,[f(Q_{j})]_{+} \equiv f (Q_{jL}) + f (Q_{jR}) &=&
f\left(Q_{j+}+ \frac{1}{2} Q_{j-}\right) +
f\left(Q_{j+}-\frac{1}{2} Q_{j-}\right).
\end{eqnarray}
Using these relations, we have
\begin{eqnarray}
\label{pmlr} (\psi^{\alpha}(x_{1}) \psi^{M\beta}(x_{1}^{'}))_{-}
&=& (\psi^{\alpha}(x_{1}) \psi^{M\beta}(x_{1}^{'}))_{L} -
(\psi^{\alpha}(x_{1}) \psi^{M\beta}(x_{1}^{'}))_{R} \nonumber \\
&=& \psi_{L}^{\alpha}(x_{1}) \psi_{L}^{M\beta}(x_{1}^{'}) -
\psi_{R}^{M\beta}(x_{1}^{'})\psi_{R}^{\alpha}(x_{1}) ,
\end{eqnarray}
substituting Eq.~(\ref{pmlr}) in Eq.~(\ref{if}), we obtain
\begin{eqnarray}
\label{if1} I  &=& \frac{1}{4!} \sum_{\eta,\kappa,\lambda,\nu =
L,R }\sum_{\alpha \neq \beta =0,1}
 \langle \hat{T}(\psi_{L}(x)
\psi_{L}^{\dagger}(\bar{x})\psi_{\eta}^{\alpha}(x_{1})
\psi_{\eta}^{M\beta}(x_{1}^{'})\psi_{\kappa}^{\alpha}(x_{2})
\psi_{\kappa}^{M\beta}(x_{2}^{'}) \nonumber \\ &&
\psi_{\lambda}^{\alpha}(x_{3}) \psi_{\lambda}^{M\beta}(x_{3}^{'})
\psi_{\nu}^{\alpha}(x_{4}) \psi_{\nu}^{M\beta}(x_{4}^{'}) )\rangle
\end{eqnarray}
Eq.~(\ref{if1}) which has $196$ terms is used to derive
Eq.~(\ref{if2}).


\section{Relations between Liouville and Hilbert Space  Green
functions}\label{ap.2pr}

In this Appendix we establish a one-to-one correspondence between
Liouville and Hilbert space Green functions.

Standard field theory is formulated in terms of  four types of
Hilbert space Green functions~\cite{kp,Keldysh,Haug}. Using the
notation of Ref.~\cite{Haug}, they are
\begin{eqnarray}
\label{cgf} G^{c}(x,\bar{x}) &\equiv& -i \langle T \psi(x)
\psi^{\dagger}(\bar{x})\rangle \nonumber \\ &=& -i \theta(t -
\bar{t})\langle \psi(x) \psi^{\dagger}(\bar{x})\rangle + i
\theta(\bar{t} - t)\langle \psi^{\dagger}(\bar{x}) \psi(x) \rangle
,
\end{eqnarray}
\begin{eqnarray}
\label{acgf} G^{\tilde{c}}(x,\bar{x}) &\equiv& -i \langle
\tilde{T} \psi(x) \psi^{\dagger}(\bar{x})\rangle \nonumber \\ &=&
-i \theta(\bar{t} - t)\langle \psi(x)
\psi^{\dagger}(\bar{x})\rangle + i \theta(t - \bar{t})\langle
\psi^{\dagger}(\bar{x}) \psi(x) \rangle ,
\end{eqnarray}
\begin{eqnarray}
\label{ggf} G^{>}(x,\bar{x}) &\equiv& -i \langle  \psi(x)
\psi^{\dagger}(\bar{x})\rangle ,
\end{eqnarray}
\begin{eqnarray}
\label{lgf} G^{<}(x,\bar{x}) &\equiv& i \langle
\psi^{\dagger}(\bar{x}) \psi(x) \rangle ,
\end{eqnarray}
$G^{c}$, $G^{\tilde{c}}$, $G^{>}$ and $G^{<}$ are known as the
time-ordered, antitime-ordered, ``greater'' and ``lesser'' Keldysh
Green functions, respectively. $T$ is a time-ordering operator in
Hilbert space, which takes any product of Fermi field operators
$\zeta$ and $\chi$ and reorders them in ascending times from right
to left, i.e.
\begin{equation}
\label{to} T \zeta(\tau_{1})\chi(\tau_{2}) \equiv
\left\{\begin{array}{ll}
 \zeta(\tau_{1})\chi(\tau_{2})& \tau_{2}<\tau_{1}\\
 -\chi(\tau_{2})\zeta(\tau_{1})& \tau_{1}<\tau_{2}\\
 \frac{1}{2}[\zeta (\tau_1) \chi (\tau_1) - \chi(\tau_1)
\zeta(\tau_1)]&\tau_{2}=\tau_{1}\end{array}\right.
\end{equation}
$\tilde{T}$ is an antitime-ordering operator in Hilbert space,
which reorders any product of Fermi field operators in descending
times from right to left, i.e.
\begin{equation}
\label{ato} \tilde{T} \zeta(\tau_{1})\chi(\tau_{2}) \equiv
\left\{\begin{array}{ll} -\chi(\tau_{2}) \zeta(\tau_{1})&
\tau_{2}<\tau_{1}\\ \zeta(\tau_{1}) \chi(\tau_{2})&
\tau_{1}<\tau_{2}\\
 \frac{1}{2}[\zeta (\tau_1) \chi (\tau_1) - \chi(\tau_1)
\zeta(\tau_1)]&\tau_{2}=\tau_{1}\end{array}\right.
\end{equation}.

Four types of the Green functions naturally show up in Liouville
space:
\begin{eqnarray}
\label{glld} G_{LL}(x,\bar{x}) = -i \langle \hat{T}
(\psi_{L}(x)\psi_{L}^{\dagger}(\bar{x}))\rangle ;
\end{eqnarray}
\begin{eqnarray}
\label{grrd} G_{RR}(x,\bar{x}) = -i \langle \hat{T}
(\psi_{R}(x)\psi_{R}^{\dagger}(\bar{x}))\rangle ;
\end{eqnarray}
\begin{eqnarray}
\label{glrd} G_{LR}(x,\bar{x}) = -i \langle \hat{T}
(\psi_{L}(x)\psi_{R}^{\dagger}(\bar{x}))\rangle ;
\end{eqnarray}
\begin{eqnarray}
\label{grld} G_{RL}(x,\bar{x}) = -i \langle \hat{T}
(\psi_{R}(x)\psi_{L}^{\dagger}(\bar{x}))\rangle,
\end{eqnarray}
where $\hat{T}$ is a time-ordering operator in Liouville space,
which rearranges all superoperators so that time decreases from
left to right, i.e.~\cite{Muk_sup}
\begin{equation}
\label{3} \hat{T} \zeta_{\nu}(\tau_{1})\chi_{\mu}(\tau_{2}) \equiv
\left\{\begin{array}{ll}
 \zeta_{\nu}(\tau_{1})\chi_{\mu}(\tau_{2})& \tau_{2}<\tau_{1}\\
 -\chi_{\mu}(\tau_{2})\zeta_{\nu}(\tau_{1})& \tau_{1}<\tau_{2}\\
 \frac{1}{2}[\zeta_{\nu} (\tau_1) \chi_{\mu} (\tau_1) - \chi_{\mu}(\tau_1)
\zeta_{\nu}(\tau_1)]&\tau_{2}=\tau_{1}\end{array}\right.
\end{equation}
where $\zeta_{\nu}(\tau)$ is a Liouville space Fermi
superoperator, and $\nu$ and $\mu$ can be either $+$, $-$, or $L$
or $R$.

We next establish the connection between the Hilbert space
(Eqs.~(\ref{cgf})-~(\ref{lgf})) and Liouville space
(Eqs.~(\ref{glld})-~(\ref{grld})) Green functions. For $G_{RL}$
and $G_{LR}$ we have
\begin{eqnarray}
\label{rlll} G_{RL}(x,\bar{x}) &=& -iTr\left[\hat{T} \psi_{R}(x)
\psi_{L}^{\dagger}(\bar{x})\rho_{0}\right]= -iTr\left[
\psi^{\dagger}(\bar{x})\rho_{0}\psi(x)\right] = -iTr\left[\psi(x)
\psi^{\dagger}(\bar{x})\rho_{0}\right] \nonumber
\\ &=& G^{>}(x,\bar{x}),
\end{eqnarray}
\begin{eqnarray}
\label{lrrr} G_{LR}(x,\bar{x}) &=& -iTr\left[\hat{T} \psi_{L}(x)
\psi_{R}^{\dagger}(\bar{x})\rho_{0}\right]= -iTr\left[\psi(x)
\rho_{0}\psi^{\dagger}(\bar{x})\right] =
-iTr\left[\psi^{\dagger}(\bar{x}) \psi(x) \rho_{0}\right]
\nonumber
\\ &=& G^{<}(x,\bar{x}),
\end{eqnarray}
where $\rho_{0}$ is unperturbed many-electron density matrix of
the isolated molecule uncoupled to the electrodes.

We next turn to $G_{RR}$ and $G_{LL}$. For $x > \bar{x}$, we have
\begin{eqnarray}
\label{rrg} G_{RR}(x,\bar{x}) &=& -iTr\left[\hat{T} \psi_{R}(x)
\psi_{R}^{\dagger}(\bar{x})\rho_{0}\right]= -iTr\left[\rho_{0}
\psi^{\dagger}(\bar{x})\psi(x)\right] \nonumber
\\ &=& -iTr\left[ \psi^{\dagger}(\bar{x})\psi(x)\rho_{0}\right],
\end{eqnarray}
\begin{eqnarray}
\label{llg} G_{LL}(x,\bar{x}) &=& -iTr\left[\hat{T} \psi_{L}(x)
\psi_{L}^{\dagger}(\bar{x})\rho_{0}\right]= -iTr\left[\psi(x)
\psi^{\dagger}(\bar{x})\rho_{0}\right].
\end{eqnarray}
For $x < \bar{x}$ we analogously obtain
\begin{eqnarray}
\label{rrl} G_{RR}(x,\bar{x}) &=& -iTr\left[\psi(x)
\psi^{\dagger}(\bar{x})\rho_{0}\right] ,
\end{eqnarray}
\begin{eqnarray}
\label{lll} G_{LL}(x,\bar{x}) &=&
-iTr\left[\psi^{\dagger}(\bar{x}) \psi(x) \rho_{0}\right] .
\end{eqnarray}
Comparing Eqs.~(\ref{rrg}),~(\ref{rrl}) with Eq.~(\ref{acgf}), we
see that $G_{RR}(x,\bar{x}) = G^{\tilde{c}}(x,\bar{x})$. Comparing
Eqs.~(\ref{llg}),~(\ref{lll}) with Eq.~(\ref{cgf}), we get
$G_{LL}(x,\bar{x}) = G^{c}(x,\bar{x})$. Eqs.~(\ref{rlll})
and~(\ref{lrrr}) provide the other relation
 $G_{RL}(x,\bar{x}) =
G^{>}(x,\bar{x})$ and $G_{LR}(x,\bar{x}) = G^{<}(x,\bar{x})$.

\section{DFT Keldysh Green functions in Hilbert space} \label{ap.3pr}

The four Hilbert space Green functions
(Eqs.~(\ref{cgf})-~(\ref{lgf})) are not independent.  To obtain
the relations between them we introduce the retarded $G^{r}$ and
the advanced $G^{a}$ Green functions\cite{Haug}
\begin{eqnarray}
\label{re} G^{r}(x,\bar{x}) &\equiv& \theta(t -
\bar{t})[G^{>}(x,\bar{x}) - G^{<}(x,\bar{x})],
\end{eqnarray}
\begin{eqnarray}
\label{ad} G^{a}(x,\bar{x}) &\equiv& \theta(\bar{t} -
t)[G^{<}(x,\bar{x}) - G^{>}(x,\bar{x})].
\end{eqnarray}

The retarded DFT Green function in real space is given by
Eq.~(3.2) of Ref.\cite{Xue}(see
Refs.~\cite{Abrikosov,Fetter,Mahan,Haug})
\begin{eqnarray}
\label{dft_green_real} G^{r}(x,\bar{x}) &=& \int \frac{d \omega}{2
\pi}e^{i\omega (\bar{t} - t)}
\sum_{\alpha}\frac{\varphi_{\alpha}(\mathbf{r})\varphi_{\alpha}^{*}(\bar{\mathbf{r}})
} {\omega - (\omega_{\alpha } - \omega_{F})  + i\delta },
\end{eqnarray}
where $\delta \longrightarrow 0$, $\omega_{\alpha}$ is the energy
of  $\alpha$ Kohn-Sham orbital molecular orbital; and $\omega_{F}$
is the Fermi energy of the molecule.

We define retarded DFT Green function $G_{\alpha }^{r}(\omega)$ of
the Kohn-Sham orbital $\alpha$ in the frequency domain
\begin{eqnarray}
\label{dft_green_defin} G^{r}(x,\bar{x}) &=& \int \frac{d
\omega}{2 \pi}e^{-i\omega (t - \bar{t})}
\sum_{\alpha}\varphi_{\alpha}(\mathbf{r})\varphi_{\alpha}^{*}(\bar{\mathbf{r}})
G_{\alpha}^{r}(\omega),
\end{eqnarray}
where
\begin{eqnarray}
\label{dft_green_orfr} G_{\alpha}^{r}(\omega) &=& \left(\omega -
\omega_{\alpha} + \omega_{F} + i\delta \right)^{-1}.
\end{eqnarray}

To connect $G^{r}$, $G^{a}$, $G^{>}$, $G^{<}$, $G^{c}$ and
$G^{\tilde{c}}$\cite{Haug,Keldysh} we use
Eqs.~(\ref{re}),~(\ref{ad}) together with
\begin{eqnarray}
\label{eq1} G^{c}(x,\bar{x}) = - G^{\tilde{c}*}(\bar{x},x),
\end{eqnarray}
\begin{eqnarray}
\label{eq2} G^{a}(x,\bar{x}) =  G^{r*}(\bar{x},x),
\end{eqnarray}
\begin{eqnarray}
\label{eq3} G^{c}(x,\bar{x}) + G^{\tilde{c}}(x,\bar{x}) =
G^{<}(x,\bar{x}) + G^{>}(x,\bar{x}),
\end{eqnarray}
Since $G^{r}(x,\bar{x})$ is given by Eq.~(\ref{dft_green_real}),
we can use the system of five equations
Eqs.~(\ref{re}),~(\ref{ad}),~(\ref{eq1})-~(\ref{eq3}) to obtain
all Keldysh DFT Green functions for the molecule alone $G^{a}$,
$G^{>}$, $G^{<}$, $G^{c}$ and $G^{\tilde{c}}$. Solving these
equations, and switching to the  Kohn-Sham orbital representation
in the frequency domain (see
Eqs.~(\ref{dft_green_real})-~(\ref{dft_green_orfr})) we obtain all
Keldysh DFT Green functions for the molecule
\begin{eqnarray}
\label{dft_green_corfr} G_{\alpha}^{c}(\omega) &=& \left(\omega +
\omega_{F} - \omega_{\alpha }  + i\delta
\right)^{-1}\theta(\omega_{F} - \omega_{\alpha }) + \left(\omega +
\omega_{F}- \omega_{\alpha } - i\delta
\right)^{-1}\theta(\omega_{\alpha } - \omega_{F}) ,
\end{eqnarray}
\begin{eqnarray}
\label{dft_green_acorfr} G_{\alpha}^{\tilde{c}}(\omega) &=& -
\left(\omega + \omega_{F} - \omega_{\alpha }  - i\delta
\right)^{-1}\theta(\omega_{F} - \omega_{\alpha }) - \left(\omega +
\omega_{F}- \omega_{\alpha } + i\delta \right)^{-1} \nonumber \\
&& \theta(\omega_{\alpha } - \omega_{F}) ,
\end{eqnarray}
\begin{eqnarray}
\label{dft_green_adcorfr} G_{\alpha}^{a}(\omega) &=& \left(\omega
- \omega_{\alpha } + \omega_{F}   - i\delta \right)^{-1},
\end{eqnarray}
\begin{eqnarray}
\label{ggff} G_{\alpha }^{>}(\omega) &=& -2\pi i (1-
\theta(\omega_{\alpha} - \omega_{F})) \delta(\omega -
\omega_{\alpha } + \omega_{F}),
\end{eqnarray}
and
\begin{eqnarray}
\label{ggff1} G_{\alpha}^{<}(\omega) &=& 2\pi i
\theta(\omega_{\alpha} - \omega_{F}) \delta(\omega -
\omega_{\alpha } + \omega_{F}).
\end{eqnarray}

The Keldysh  Green functions for the electrode $A$ can be obtained
analogously by replacing $G^{r}(x,\bar{x})$  with
$D_{A}^{r}(x,\bar{x})$.
 The retarded Green function of an electron
$D_{A}^{r}(\omega)$ in a metal electrode in frequency domain is
\begin{eqnarray}
\label{metgf} D_{A}^{r}(\omega) \simeq (\omega - \mu_{A} +
i\delta)^{-1},
\end{eqnarray}
where $\mu_{A}$ is the Fermi energy level of electrons in
electrode $A$ ($B$). The difference between the chemical
potentials of the electrodes $A$ and $B$ is given by the external
voltage $U = \mu_{A} - \mu_{B}$. In Eq.~(\ref{metgf}) we assumed
that the electrons in the metal are at the Fermi level.
\begin{eqnarray}
\label{dft_green_corfrm} D_{A}^{c}(\omega) &=& \left(\omega -
\mu_{A} + i\delta \right)^{-1}\theta(\omega - \mu_{A}) +
\left(\omega - \mu_{A}  - i\delta \right)^{-1} \nonumber \\ &&
\theta(\mu_{A} - \omega ) ,
\end{eqnarray}
\begin{eqnarray}
\label{dft_green_acorfrm} D_{A}^{\tilde{c}}(\omega) &=&
-\left(\omega - \mu_{A} - i\delta \right)^{-1}\theta(\omega -
\mu_{A}) - \left(\omega - \mu_{A}  + i\delta \right)^{-1}
\nonumber \\ && \theta(\mu_{A} - \omega ) ,
\end{eqnarray}
\begin{eqnarray}
\label{dft_green_adcorfrm} D_{A}^{a}(\omega) &=& \left(\omega -
\mu_{A} - i\delta \right)^{-1},
\end{eqnarray}
\begin{eqnarray}
\label{ggffm} D_{A}^{>}(\omega) &=& -2\pi i (1 -  \delta(\omega -
\mu_{A})),
\end{eqnarray}
\begin{eqnarray}
\label{ggff1m} D_{A}^{<}(\omega) &=& 2\pi i   \delta(\omega -
\mu_{A}).
\end{eqnarray}
Similar expressions apply to electrode $B$. The final expression
for the molecular current only depends on the sum of the Green
functions of electrodes $A$ and $B$ ($D^{c} = D_{A}^{c} +
D_{B}^{c}$, $D^{\tilde{c}} = D_{A}^{\tilde{c}} +
D_{B}^{\tilde{c}}$, $D^{>} = D_{A}^{>} + D_{B}^{>}$ and $D^{<} =
D_{A}^{<} + D_{B}^{<}$).

\section{The Various Contributions to the Molecular Current} \label{ap.4pr}

Only terms which contain the contractions $\langle \hat{T}
(\psi_{L}(x_{n})\psi_{L}^{\dagger}(x_{m}))\rangle$, $\langle
\hat{T} (\psi_{R}(x_{n})\psi_{R}^{\dagger}(x_{m}))\rangle$,
$\langle \hat{T}
(\psi_{R}(x_{n})\psi_{L}^{\dagger}(x_{m}))\rangle$, and $\langle
\hat{T} (\psi_{L}(x_{n})\psi_{R}^{\dagger}(x_{m}))\rangle$ (and
similar contractions for the electrodes) can contribute to $I$ in
Eq.~(\ref{if2}), because all other contractions vanish. Therefore,
only twenty one terms survive in the r.h.s. of Eq.~(\ref{if2}).
These terms contain the four types of Liouville space Green
functions of the molecule $G_{LL}$, $G_{RR}$, $G_{LR}$, and
$G_{RL}$ introduced in Appendix~\ref{ap.2pr} and of the electrodes
$D_{LL} = D_{LL(A)} + D_{LL(B)}$, $D_{RR} = D_{RR(A)} +
D_{RR(B)}$, $D_{LR} = D_{LR(A)} + D_{LR(B)}$, and $D_{RL} =
D_{RL(A)} + D_{RL(B)}$.

 As an illustration we write the expression for one of
the terms contributing to $S$,
\begin{eqnarray}
\label{i15} I_{15} &=& -i
U(x_{1},x_{1}')U(x_{2},x_{2}')U(x_{3},x_{3}')U(x_{4},x_{4}')G_{RL}(x_{1},x_{2})G_{LR}
(x_{3},x_{4})G_{LL}(x,\bar{x}) \nonumber\\ &&
D_{LR}(x_{2}',x_{1}')D_{RL}(x_{4}',x_{3}').
\end{eqnarray}
Using the relations of Appendix \ref{ap.2pr}, we can recast this
in the form of Eq.~(\ref{it3}).

 Below we present all twenty one
terms which contribute to the current (Eq.~(\ref{current_res})),
using the Kohn-Sham orbitals representation in the frequency
domain. The Greek indices run over the occupied Kohn-Sham orbitals
$\alpha, \beta, \gamma = 1 \ldots N$.
\begin{eqnarray}
\label{ie1}I_{1} &=& - i \sum_{\alpha} U_{\beta}U_{\alpha}^{3}
G_{\beta}^{>}(\omega)
G_{\alpha}^{c}(\omega_{1})G_{\alpha}^{\tilde{c}}(\omega_{2})D^{\tilde{c}*}(-\omega_{2
})D^{<}(-\omega_{1});
\end{eqnarray}
\begin{eqnarray}
\label{ie2}I_{2} &=& - i \sum_{\alpha} U_{\beta}U_{\alpha}^{3}
G_{\beta}^{>}(\omega)
G_{\alpha}^{c}(\omega_{1})G_{\alpha}^{c}(\omega_{2})D^{<}(-\omega_{2})D^{c*}(-\omega_{1});
\end{eqnarray}
\begin{eqnarray}
\label{ie3}I_{3} &=& - i \sum_{\alpha} U_{\beta}^{3}U_{\alpha }
G_{\beta }^{>}(\omega) G_{\alpha
}^{<}(\omega_{1})G_{\beta}^{>}(\omega_{2})D^{\tilde{c}*}(-\omega_{2})D^{<}(-
\omega_{1});
\end{eqnarray}
\begin{eqnarray}
\label{ie4}I_{4} &=&  i \sum_{\alpha} U_{\beta}^{2}U_{\alpha }^{2}
G_{\beta}^{c}(\omega_{2})
G_{\alpha}^{c}(\omega)G_{\alpha}^{>}(\omega_{1})D^{c*}(-\omega_{2})D^{<}(-\omega_{1});
\end{eqnarray}
\begin{eqnarray}
\label{ie5}I_{5} &=&  - i \sum_{\alpha} U_{\beta}U_{\alpha }^{3}
G_{\beta}^{c}(\omega_{1}) G_{\alpha
}^{>}(\omega)G_{\alpha}^{c}(\omega_{2})D^{<}(-\omega_{2})D^{c*}(-\omega_{1});
\end{eqnarray}
\begin{eqnarray}
\label{ie6}I_{6} &=&  - i \sum_{\alpha} U_{\beta}^{2}U_{\alpha
}^{2} G_{\beta}^{c}(\omega_{1}) G_{\alpha
}^{>}(\omega)G_{\alpha}^{\tilde{c}}(\omega_{2})D^{<}(-\omega_{2})D^{\tilde{c}*}
(-\omega_{1}).
\end{eqnarray}
\begin{eqnarray}
\label{ih1}I_{7} &=& - i \sum_{\alpha} U_{\beta}U_{\alpha}^{3}
G_{\beta}^{<}(\omega)
G_{\alpha}^{c}(\omega_{1})G_{\alpha}^{\tilde{c}}(\omega_{2})D^{\tilde{c}*}(-\omega_{2
})D^{>}(-\omega_{1});
\end{eqnarray}
\begin{eqnarray}
\label{ih2}I_{8} &=& - i \sum_{\alpha} U_{\beta}^{3}U_{\alpha }
G_{\beta }^{<}(\omega) G_{\alpha
}^{c}(\omega_{1})G_{\beta}^{c}(\omega_{2})D^{>}(-\omega_{2})D^{*}(-\omega_{1});
\end{eqnarray}
\begin{eqnarray}
\label{ih3}I_{9} &=& - i \sum_{\alpha} U_{\beta}U_{\alpha}^{3}
G_{\beta}^{<}(\omega)
G_{\alpha}^{>}(\omega_{1})G_{\alpha}^{<}(\omega_{2})D^{*}(-\omega_{2})D^{>}(
-\omega_{1});
\end{eqnarray}
\begin{eqnarray}
\label{ih4}I_{10} &=&  i \sum_{\alpha} U_{\beta}^{2}U_{\alpha
}^{2} G_{\beta }^{c}(\omega_{2}) G_{\alpha
}^{c}(\omega)G_{\alpha}^{<}(\omega_{1})D^{c*}(-\omega_{2})D^{>}(-\omega_{1});
\end{eqnarray}
\begin{eqnarray}
\label{ih5}I_{11} &=&  - i \sum_{\alpha} U_{\beta}U_{\alpha }^{3}
G_{\beta }^{c}(\omega_{1}) G_{\alpha
}^{<}(\omega)G_{\alpha}^{c}(\omega_{2})D^{>}(-\omega_{2})D^{c*}(-\omega_{1});
\end{eqnarray}
\begin{eqnarray}
\label{ih6}I_{12} &=&  - i \sum_{\alpha} U_{\beta}^{2}U_{\alpha
}^{2} G_{\beta }^{c}(\omega_{1}) G_{\alpha
}^{<}(\omega)G_{\alpha}^{\tilde{c}}(\omega_{2})
D^{>}(-\omega_{2})D^{\tilde{c}*}(-\omega_{1})  .
\end{eqnarray}
\begin{eqnarray}
\label{it1}I_{13} &=& - i  \sum_{\alpha
\gamma}U_{\beta}U_{\gamma}U_{\alpha }^{2} G_{\beta}^{c}(\omega)
G_{\alpha}^{c}(\omega_{1})G_{\gamma}^{c}(\omega_{2})D^{c*}(-\omega_{2})D^{c*}(-\omega
_{1});
\end{eqnarray}
\begin{eqnarray}
 \label{it2}I_{14} &=& - i \sum_{\alpha \gamma}
U_{\beta}U_{\gamma}U_{\alpha }^{2} G_{\beta}^{c}(\omega)
G_{\alpha}^{c}(\omega_{1})G_{\gamma}^{\tilde{c}}(\omega_{2})D^{c*}(-\omega_{1})D^{\tilde{c}*}(-\omega_{2});
\end{eqnarray}
\begin{eqnarray}
\label{it3}I_{15} &=& - i \sum_{\alpha \gamma}
U_{\beta}U_{\gamma}U_{\alpha }^{2} G_{\beta}^{c}(\omega)
G_{\alpha}^{<}(\omega_{1})G_{\gamma}^{>}(\omega_{2})
D^{<}(-\omega_{2})D^{>}(-\omega_{1})  ;
\end{eqnarray}
\begin{eqnarray}
\label{it4}I_{16} &=& - i \sum_{\alpha \gamma}
U_{\beta}U_{\gamma}U_{\alpha }^{2}
G_{\beta}^{\tilde{c}}(\omega_{2}) G_{\alpha
}^{<}(\omega_{1})G_{\gamma}^{>}(\omega)D^{\tilde{c}*}(-\omega_{2})D^{>}(-\omega_{1});
\end{eqnarray}
\begin{eqnarray}
\label{it5}I_{17} &=& i \sum_{\alpha \gamma}
U_{\beta}U_{\gamma}U_{\alpha }^{2} G_{\beta}^{<}(\omega_{1})
G_{\alpha
}^{\tilde{c}}(\omega)G_{\gamma}^{>}(\omega_{2})D^{<}(-\omega_{2})D^{>}(-\omega_{1});
\end{eqnarray}
\begin{eqnarray}
\label{it6}I_{18} &=&  i \sum_{\alpha \gamma}
U_{\beta}U_{\gamma}U_{\alpha }^{2}
G_{\beta}^{\tilde{c}}(\omega_{1}) G_{\alpha
}^{\tilde{c}}(\omega_{2})G_{\gamma}^{c}(\omega)D^{\tilde{c}*}(-\omega_{2})D^{\tilde{c
}*}(-\omega_{1});
\end{eqnarray}
\begin{eqnarray}
\label{it7}I_{19} &=& - i \sum_{\alpha \gamma} U_{\beta}U_{\gamma
}U_{\alpha }^{2} G_{\beta}^{<}(\omega)
G_{\alpha}^{>}(\omega_{2})G_{\gamma}^{c}(\omega_{1})D^{<}(-\omega_{2})D^{c*}(-\omega_{1});
\end{eqnarray}
\begin{eqnarray}
\label{it8}I_{20} &=& - i \sum_{\alpha \gamma} U_{\beta}U_{\gamma
}U_{\alpha }^{2} G_{\beta}^{\tilde{c}}(\omega_{1})
G_{\alpha}^{>}(\omega_{2})G_{\gamma}^{<}(\omega)D^{<}(-\omega_{2})D^{\tilde{
c}*}(-\omega_{1}) ;
\end{eqnarray}
\begin{eqnarray}
\label{it9}I_{21} &=& - i \sum_{\alpha \gamma}
U_{\beta}U_{\gamma}U_{\alpha }^{2} G_{\beta}^{<}(\omega) G_{\alpha
}^{c}(\omega_{2})G_{\gamma}^{>}(\omega_{1})D^{c*}(-\omega_{2})D^{<}(-\omega_{1}
).
\end{eqnarray}

\section{Relation between Boson Green functions in Liouville and Hilbert
Space}\label{ap.5pr}

The relations between Liouville space Green functions and Keldysh
Hilbert space Green functions for fermions also apply to  boson
 fields. We consider a system of identical atoms in
an external potential~\cite{ChernyakChoi}. The definitions for the
Liouville space Green functions $G_{LL}$, $G_{LR}$, $G_{RL}$ and
$G_{RR}$ (Eqs.~(\ref{glld})-~(\ref{grld})) are the same for bosons
as for fermions with the only difference that a time-ordering
operator in Liouville space $\hat{T}_{B}$ is defined
as~\cite{Muk_sup}
\begin{equation}
\label{31} \hat{T}_{B} \zeta_{\nu}(\tau_{1})\chi_{\mu}(\tau_{2})
\equiv \left\{\begin{array}{ll}
 \zeta_{\nu}(\tau_{1})\chi_{\mu}(\tau_{2})& \tau_{2}<\tau_{1}\\
 \chi_{\mu}(\tau_{2})\zeta_{\nu}(\tau_{1})& \tau_{1}<\tau_{2}\\
 \frac{1}{2}[\zeta_{\nu} (\tau_1) \chi_{\mu} (\tau_1) + \chi_{\mu}(\tau_1)
\zeta_{\nu}(\tau_1)]&\tau_{2}=\tau_{1}\end{array}\right.
\end{equation}
where $\zeta_{\nu}(\tau)$ is a Liouville space Bose superoperator,
and $\nu$ and $\mu$ can be either $+$, $-$, or $L$ or $R$.

The four types of Hilbert space Green functions for bosons are
defined as~\cite{Keldysh}
\begin{eqnarray}
\label{cgf1} G^{c}(x_{1},x_{2}) &\equiv& -i \langle T_{B}
\psi(x_{1}) \psi^{\dagger}(x_{2})\rangle \nonumber \\ &=& -i
\theta(t_{1} - t_{2})\langle \psi(x_{1})
\psi^{\dagger}(\bar{x})\rangle - i \theta(t_{2} - t_{1})\langle
\psi^{\dagger}(x_{2})  \psi(x_{1}) \rangle ,
\end{eqnarray}
\begin{eqnarray}
\label{acgf1} G^{\tilde{c}}(x,\bar{x}) &\equiv& -i \langle
\tilde{T}_{B} \psi(x) \psi^{\dagger}(\bar{x})\rangle \nonumber \\
&=& -i \theta(t_{6} - t_{5})\langle \psi(x)
\psi^{\dagger}(\bar{x})\rangle - i \theta(t_{5} - t_{6})\langle
\psi^{\dagger}(\bar{x}) \psi(x) \rangle ,
\end{eqnarray}
\begin{eqnarray}
\label{ggf1} G^{>}(x_{1},x_{2}) &\equiv& -i \langle  \psi(x_{1})
\psi^{\dagger}(x_{2})\rangle ,
\end{eqnarray}
\begin{eqnarray}
\label{lgf1} G^{<}(x_{1},x_{2}) &\equiv& - i \langle
\psi^{\dagger}(x_{2}) \psi(x_{1}) \rangle .
\end{eqnarray}

In Eq.~(\ref{cgf1}) $T_{B}$ is a time-ordering operator in Hilbert
space for bosons, which takes any product of Bose field operators
$\zeta$, $\chi$ and reorders them in ascending times from right to
left, i.e.
\begin{equation}
\label{to1} T_{B} \zeta(\tau_{1})\chi(\tau_{2}) \equiv
\left\{\begin{array}{ll}
 \zeta(\tau_{1})\chi(\tau_{2})& \tau_{2}<\tau_{1}\\
 \chi(\tau_{2})\zeta(\tau_{1})& \tau_{1}<\tau_{2}\\
 \frac{1}{2}[\zeta (\tau_1) \chi (\tau_1) + \chi(\tau_1)
\zeta(\tau_1)]&\tau_{2}=\tau_{1}\end{array}\right.
\end{equation}

$\tilde{T}_{B}$ is an antitime-ordering operator in Hilbert space
for bosons, which reorders any product of Bose field operators in
descending times from right to left, i.e.
\begin{equation}
\label{ato1} \tilde{T}_{B} \zeta(\tau_{1})\chi(\tau_{2}) \equiv
\left\{\begin{array}{ll} \chi(\tau_{2}) \zeta(\tau_{1})&
\tau_{2}<\tau_{1}\\ \zeta(\tau_{1}) \chi(\tau_{2})&
\tau_{1}<\tau_{2}\\
 \frac{1}{2}[\zeta (\tau_1) \chi (\tau_1) + \chi(\tau_1)
\zeta(\tau_1)]&\tau_{2}=\tau_{1}\end{array}\right.
\end{equation}

In analogy with Appendix \ref{ap.2pr}, we can show that for bosons
$G_{LL}(x_{1},x_{2}) = G^{c}(x_{1},x_{2})$, $G_{RR}(x_{1},x_{2}) =
G^{\tilde{c}}(x_{1},x_{2})$, $G_{RL}(x_{1},x_{2}) =
G^{>}(x_{1},x_{2})$, $G_{LR}(x_{1},x_{2}) = G^{<}(x_{1},x_{2})$.

The retarded boson Green function in real space is given
by~\cite{Keldysh}
\begin{eqnarray}
\label{dft_green_real1} G^{r}(x_{1},x_{2}) &=& \int \frac{d
\omega}{2 \pi}e^{i\omega (t_{2} - t_{1})}
\sum_{\alpha}\frac{\varphi_{\alpha}(\mathbf{r}_{1})\varphi_{\alpha}^{*}(\mathbf{r}_{2
})} {\omega - (\omega _{\alpha } - \mu)  + i\delta },
\end{eqnarray}
where $\delta \longrightarrow 0$; $\omega_{\alpha}$ is the energy
of  $\alpha$ eigenvalue; $\mu$ is the chemical potential;
$\varphi_{\alpha}(\mathbf{r}_{1})$ is the wavefunction
corresponding to $\alpha$ eigenvalue of the external potential.

We define retarded boson Green function $G_{\alpha }^{r}(\omega)$
in the frequency domain
\begin{eqnarray}
\label{dft_green_defin1} G^{r}(x_{1},x_{2}) &=& \int \frac{d
\omega}{2 \pi}e^{-i\omega (t_{1} - t_{2})}
\sum_{\alpha}\varphi_{\alpha}(\mathbf{r}_{1})\varphi_{\alpha}^{*}(\mathbf{r}_{2})
G_{\alpha}^{r}(\omega),
\end{eqnarray}
where
\begin{eqnarray}
\label{dft_green_orfr1} G_{\alpha}^{r}(\omega) &=& \left(\omega -
\omega_{\alpha} + \mu + i\delta \right)^{-1}.
\end{eqnarray}

The Keldysh Green functions for bosons are~\cite{Keldysh}
\begin{eqnarray}
\label{dft_green_corfr1} G_{\alpha}^{c}(\omega) &=& \left(\omega +
\mu - \omega_{\alpha }  + i\delta \right)^{-1} - 2\pi i
n_{\alpha}\delta(\omega - \omega_{\alpha} + \mu)  ,
\end{eqnarray}
\begin{eqnarray}
\label{dft_green_acorfr1} G_{\alpha}^{\tilde{c}}(\omega) &=& -
\left(\omega + \mu - \omega_{\alpha } - i\delta \right)^{-1} -
2\pi i n_{\alpha}\delta(\omega - \omega_{\alpha} + \mu)  ,
\end{eqnarray}
\begin{eqnarray}
\label{dft_green_adcorfr1} G_{\alpha}^{a}(\omega) &=& \left(\omega
- \omega_{\alpha } + \mu)   - i\delta \right)^{-1},
\end{eqnarray}
\begin{eqnarray}
\label{ggff11} G_{\alpha }^{>}(\omega) &=& -2\pi i (1+n_{\alpha})
\delta(\omega - \omega_{\alpha } + \mu),
\end{eqnarray}
\begin{eqnarray}
\label{lgff1} G_{\alpha}^{<}(\omega) &=& - 2\pi i n_{\alpha}
\delta(\omega - \omega_{\alpha } + \mu),
\end{eqnarray}
where $n_{\alpha} = (\exp[(\omega_{\alpha } - \mu)/(k_{B}T)] -
1)^{-1}$ is the occupation number of the $\alpha$ harmonic
oscillator ($T$ is temperature, and $k_{B}$ is Boltzmann
constant).


\end{document}